\documentclass[journal]{IEEEtran}

\ifCLASSINFOpdf
\else
   \usepackage[dvips]{graphicx}
\fi
\usepackage{url}

\usepackage{graphicx}

\usepackage{pifont}
\newcommand{\cmark}{\ding{51}}

\usepackage{multirow}
\usepackage{amsfonts}

\usepackage{algorithm}
\usepackage{algorithmic}

\usepackage{comment}

\usepackage[table,xcdraw]{xcolor}

\usepackage{color, colortbl}
\definecolor{Gray}{gray}{0.9}

\usepackage{mathtools}
\usepackage{amsmath}

\usepackage{hyperref}

\usepackage{svg}
\newcommand{\orcid}[1]{\href{https://orcid.org/#1}{\includegraphics[width=10pt]{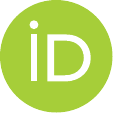}}}

\begin{document}

\title{Parameter-Free Logit Distillation via Sorting Mechanism}

\author{
Stephen Ekaputra Limantoro \orcid{0000-0002-1993-5303}

\IEEEcompsocitemizethanks{\IEEEcompsocthanksitem
The author is with the Department of Electrical Engineering and Computer Science, National Yang Ming Chiao Tung University, Hsinchu  30010, Taiwan. Email: (stephen.ee08@nycu.edu.tw).
}
}

\maketitle

\begin{abstract}
Knowledge distillation (KD) aims to distill the knowledge from the teacher (larger) to the student (smaller) model via soft-label for the efficient neural network. In general, the performance of a model is determined by accuracy, which is measured with labels. However, existing KD approaches usually use the teacher with its original distribution, neglecting the potential of incorrect prediction. This may contradict the motivation of hard-label learning through cross-entropy loss, which may lead to sub-optimal knowledge distillation on certain samples. To address this issue, we propose a novel logit processing scheme via a sorting mechanism. Specifically, our method has a two-fold goal: (1) fixing the incorrect prediction of the teacher based on the labels and (2) reordering the distribution in a natural way according to priority rank at once. As an easy-to-use, plug-and-play pre-processing, our sort method can be effectively applied to existing logit-based KD methods. Extensive experiments on the CIFAR-100 and ImageNet datasets demonstrate the effectiveness of our method. 
\end{abstract}


\IEEEpeerreviewmaketitle

\section{Introduction}
\IEEEPARstart{O}{ver} the past decade, the emergence of deep neural networks (DNNs) has transformed the field of computer vision tasks. The advancement of the DNN is associated with an increase in model size, demonstrating that larger models often yield better performance. To tackle this issue, knowledge distillation (KD) \cite{modelcompression,firstkd,kd} was introduced to cut down the model size and capacity. Specifically, KD aids in the training of small student networks through the knowledge of larger pre-trained teacher networks. This technique can effectively improve the student networks without any additional computation cost.

\begin{figure}[t]
\centering
\includegraphics[scale=0.38]{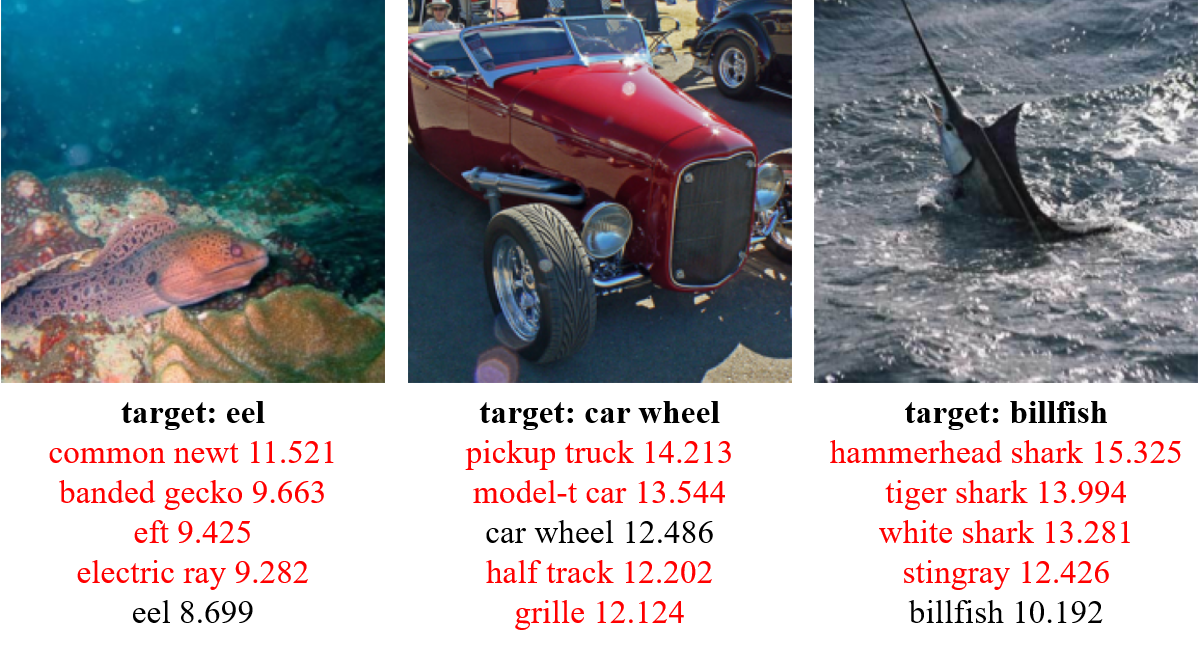}
\caption{Visualizations of incorrect prediction with top-5 values on ImageNet. We take the ResNet50 model as the pre-trained teacher model.}
\vspace{-0.5em}
\label{fig:problem}
\end{figure}

Most of the existing logit-based KD methods \cite{kd,ctkd,lskd} directly transfer the knowledge from the pre-trained teacher networks to the student networks via soft labels. Even though the teacher networks generally demonstrate superior performance on designated tasks, it remains a possibility that the prediction is not always accurate. The situation arises when the target confidence is not the highest. Fig. \ref{fig:problem} shows the incorrect prediction case of the teacher model. In the top-5 predictions, we can observe that the predicted categories share highly correlated features and semantics, which are prone to misclassification. For instance, car wheels and model-t cars belong to the category of vehicles with wheels. Therefore, relying on these false predictions instead of the ground truth for guidance may result in students performing sub-optimally on misclassified samples. 

The utilization of labels is essential for guidance to ensure that the teacher's prediction is correct. The advanced use of labels has been studied previously through regularization or self-training \cite{lsr,tfkd} and logit transformation \cite{bettersupervision,swap,lskd,improvekd} to improve the network performance. Label smooth regularization (LSR) \cite{lsr} uses one-hot labels to generate fixed smooth soft labels. Unlike LSR, in which domain teacher models are unused, our approach utilizes label information to modify the teacher's output in KD. A swap mechanism is introduced to fix the incorrect prediction of the teachers via labels through logit with temperature adjustments \cite{bettersupervision} and bi-level teachers \cite{swap}. Specifically, it swaps the target confidence with the misclassified prediction. However, the swap mechanism will devalue the high-correlated semantic confidence if the target confidence is far below that of other top non-target confidences. LSKD \cite{lskd} utilizes z-score normalization for the logits, ignoring the label usage on logit transformation. Recently, LDA \cite{improvekd} balances between the teacher's confidence and the target label through a weighted mix. Different from ours, LDA adjusts the entire probability when the prediction is false. In this work, our approach draws parallels to the underlying principles of the swap method by pre-processing the teacher's logit and recycling the existing confidence.

\begin{figure*}[ht]
\centering
\includegraphics[scale=0.61]{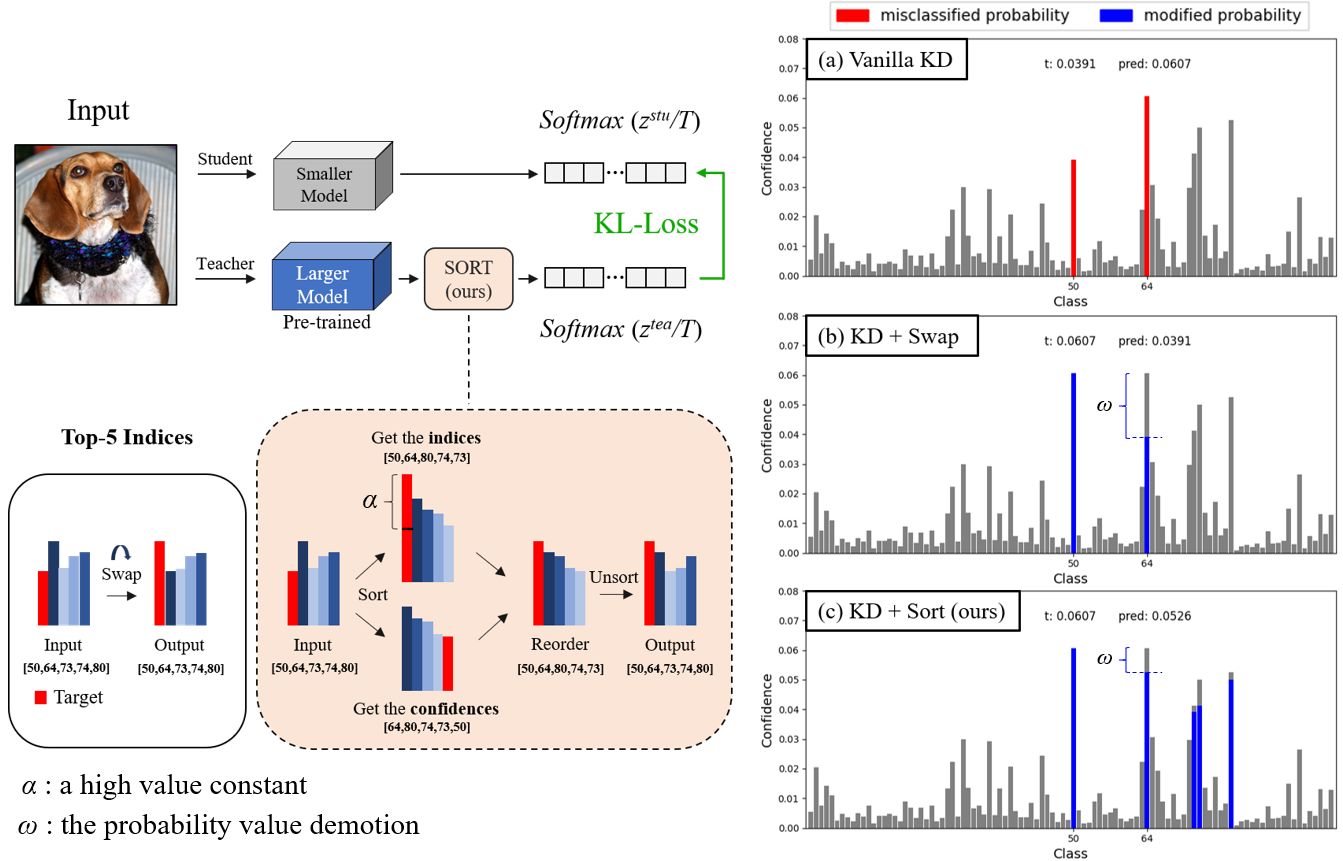}
\caption{Overview of Sort-KD. (a) Classical KD directly uses the teacher's original distribution. (b) The swap mechanism \cite{swap} is introduced to fix the teacher's incorrect prediction by swapping the target index with the non-target with the highest probability, which only affects two indices. (c) The proposed sorting mechanism mitigates the side-effect of the swap mechanism when the target index of the original distribution is not in the top-2. Concretely, the swapped non-target confidence drastically gets demoted by $\omega$ even though it still contains a useful context. We can see the confidence of the 64th class index in (b)\&(c) is different. Using a real sample on CIFAR-100, we show that the sorting mechanism affects the top-5 distribution and reorders the distribution more naturally.}
\label{fig:framework}
\end{figure*}

To this end, we introduce a novel parameter-free technique to transform the teacher's prediction based on the true label. Specifically, the misclassified target prediction is corrected, while the other non-target ones are reordered. On the other hand, the correct prediction is skipped. Through our approach, the probability distribution is still natural since the sorting mechanism explicitly recycles the existing confidences without introducing new values. Compared to the swap method \cite{swap}, our sorting mechanism eliminates the potential issue of significant devaluation of a high-correlated non-target. In this work, we apply our method to existing logit-based KD methods as a plug-and-play pre-processing and demonstrate its effectiveness in negligible costs.

The summaries of our main contributions are as follows:
\begin{itemize}
\item We expose the shortcomings of classical KD and the swap method regarding naturality and robustness. This prevents students from acquiring accurate semantics from the teachers on misclassified samples.
\item We propose a novel logit processing scheme via a sorting mechanism to cope with the false prediction of the teacher models based on labels. By sorting mechanisms, the prediction will be correct, and the non-target indices with higher confidence will be effectively reordered without additional parameters.
\item We present extensive experiments with various teacher and student models on the CIFAR-100 and ImageNet datasets. We demonstrate the effectiveness of our method as a plug-and-play pre-processing on the teacher's logit output.
\end{itemize}

\section{Related Work}
KD \cite{kd} aims to transfer the knowledge from a pre-trained teacher model to a small student model via soft labels. Learning from the soft labels provided by a teacher enables students to attain improved performance compared to training solely on complex labels. The knowledge transfer is done by minimizing the divergence between predictions from the teacher and student models. Generally, KD as representative works can be classified into two types, \textit{i.e.}, logit-based \cite{crd,dkd,ctkd,mlkd,lskd} and feature-based \cite{kdab,ofd,reviewkd} knowledge distillation. In this work, we exploit logit-based KD methods to demonstrate our proposed pre-processing method.

Previous existing logit-based distillation pipelines \cite{kd,dkd,ctkd,lskd} mainly focus on the use of the teacher's predictions to distill the knowledge to the student model. While the teacher model provides valuable insights, it is important to note that its predictions are not always accurate. This inaccuracy can potentially lead a student model to deficient outcomes. SLD \cite{swap} effectively solves this issue by swapping the misclassified prediction value with the logit maximum value to fix the correctness of the teacher's prediction. Nevertheless, we argue that the swap method can demote meaningful probability, which is still correlated with the target. To overcome this issue, we modify the swap method by sorting the distribution to eliminate its side effects. As a result, the distribution will be smoother.

\section{Sort Knowledge Distillation}
In this section, we first start with the preliminary. We then describe the details of the proposed method's sorting mechanism and discuss the advantages of our proposed method.

\subsection{Preliminary}
The notion of knowledge distillation is initially proposed by \cite{kd}. Given the labeled dataset $D = \{(x,y)\}$ as an input, student $f^{stu}$ and teacher $f^{tea}$ models respectively predict logit vector $z$. Therefore, $z^{stu} = f^{stu}(x)$ and $z^{tea} = f^{tea}(x)$. The prediction output ${z\in \mathbb{R}^{C}}$ with $C$ number of classes is processed with the softmax function into probability:

\begin{equation}
\begin{aligned}
\label{softmax}
p_{j} = \frac{exp{(z_{j}/T)}}{\sum^{C}_{c=1} exp{(z_{c}/T)}},
\end{aligned}
\end{equation}

where $z_{j}$ and $p_{j}$ represents the logit and probability output on the $j$-th class, respectively. $T$ is the temperature scaling to adjust the softness of the distribution. Then, the Kullback-Leibler divergence loss (KL) is used to minimize the discrepancy between the student and teacher probability output:

\begin{equation}
\begin{aligned}
\label{kdloss}
\mathcal{L}_{KD} = KL(p^{tea} || p^{stu}) = \sum^{C}_{j=1} p_{j}^{tea} log \left( \frac{p_{j}^{tea}} {p_{j}^{stu}} \right), 
\end{aligned}
\end{equation}

where $\mathcal{L}_{KD}$ is the KL loss, $p_{j}^{tea}$ and $p_{j}^{stu}$ are the teacher's and student's probability output on the $j$-th class, respectively.

\subsection{Sorted Teacher Logit}
As shown in Fig. \ref{fig:framework}, we propose a sorting mechanism for the teacher's output to improve the classical KD and demonstrate the comparison with the swap mechanism \cite{swap}.

We first create a modified logit of the teacher by adding the target with a high-value constant before sorting it to ensure that the target index is the highest. This can be written as: 

\begin{equation}
\begin{aligned}
\label{mask}
    M_{j} = 
\begin{cases}
    \alpha,& \text{if } z^{tea}_{j} = y\\
    0, & \text{otherwise}
\end{cases}
\end{aligned}
\end{equation}

\begin{equation}
\begin{aligned}
\label{modified_z}
z^{tea'} = z^{tea} + M,
\end{aligned}
\end{equation}

where $y$ is the target, $M$ is the one-hot target, and $\alpha$ is a high value constant. We set $\alpha > max(z^{tea})$. 

After obtaining the modified teacher's logit, it is sorted in descending order. The objective is to generate the expected indices order, as the first index of the sorted indices is the target one. On the other hand, we also sort the original teacher's logit in descending order to get the unmodified or original distribution. These can be written as:

\begin{equation}
\begin{aligned}
\label{sort1}
z^{temp'} = z^{tea'}_{\phi(j)}, I^{temp'} = \phi(j),
\end{aligned}
\end{equation}

\begin{equation}
\begin{aligned}
\label{sort2}
z^{temp} = z^{tea}_{\phi(j)}, I^{temp} = \phi(j),
\end{aligned}
\end{equation}

where $z^{temp'}$ and $I^{temp'}$ denote sorted prediction and indices of modified logit, respectively, from Eq. \ref{modified_z}. On the other hand, $z^{temp}$ and $I^{temp}$ denote sorted prediction and indices of the original teacher logit, respectively. $\phi$ is a permutation of indices $j \in \{0,1,\dots, C-1\}$, sorted in descending order.

By taking the expected indices $I^{temp'}$ and sorted original distribution $z^{temp}$, we can transform the $z^{temp}$ to new sorted teacher $z^{sorted\_tea}$ with $I^{temp'}$. It is presented as:

\begin{equation}
\begin{aligned}
\label{new_z}
z^{sorted\_tea}_{j} = z^{temp}_{I^{temp'}(j)},
\end{aligned}
\end{equation}

Finally, we have designed the new sorted teacher. We further minimize the discrepancy between sorted teacher and student models with the KL-divergence loss in the following:

\begin{equation}
\begin{aligned}
\label{sort_loss}
\mathcal{L}_{Sort-KD} = KL(p^{sorted\_tea} || p^{stu}),
\end{aligned}
\end{equation}

where $\mathcal{L}_{Sort-KD}$ is the designed loss and $z^{sorted\_tea}$ presents the new sorted teacher.

\subsection{Discussion}
In this subsection, we describe the fundamental reasons that make the sorting mechanism natural and robust in modifying the teacher's predictions.

\noindent\textbf{Natural.} The sorting mechanism is natural because the distribution is from the original prediction output. Concretely, the sum of the sorted prediction's output is equivalent to the sum of the original and swapped \cite{swap} output. Henceforth, the relationship between the temperature and the softmax function in KD remains consistent, regardless of how the probability's smoothness changes. We can see in the following:

\begin{equation}
\begin{aligned}
\label{sort_equal_loss}
\sum^{C}_{j=1} z_{j}^{tea} = \sum^{C}_{j=1} z_{j}^{swapped\_tea} = \sum^{C}_{j=1} z_{j}^{sorted\_tea},
\end{aligned}
\end{equation}

\noindent\textbf{Robust.} Notably, it is guaranteed that the prediction of the teacher is 100\% correct. Compared with the swap mechanism, the advantage of our proposed method comes from when the confidence of the target index is top-(2+$n$) where $n \in \mathbb{Z}, n>0 $. In this case, if we use a swap mechanism, we first need to obtain the top-$k$ rank of the target and swap the index multiple times to be the same as our proposed method's output. For example, given the target confidence is top-3, we need to swap it with top-2 and do it again with top-1. Otherwise, the confidence of the swapped index will be devalued drastically, even though the confidence still contains useful semantics. With the proposed sorting mechanism, no matter what the top-$k$ is, the output is reordered in a smoother way than the swap mechanism based on labels and original distribution at once.

\section{Experiment Results}
\subsection{Dataset}
We conduct experiments on CIFAR-100 \cite{cifar}, and ImageNet \cite{imagenet} datasets for the image classification tasks. CIFAR-100 is a well-known image classification dataset with a resolution of 32x32 pixels and 100 categories, consisting of 50,000 training and 10,000 validation images. ImageNet, a large resolution dataset, is one of the most important benchmark datasets for image classification and contains around 1.3 million training and 50,000 validation images.

\subsection{Model Setup}
We conduct experiments with various architectures, including ResNets \cite{resnet}, WRNs \cite{wideresnet}, VGGs \cite{vgg}, ShuffleNets \cite{shufflenet}, and MobileNets \cite{mobilenet,mobilenetv2}. We perform all experiments for teacher-student pairs in two settings, \textit{i.e.,} identical structures and distinct architecture structures.

\begin{table*}[ht]
\renewcommand{\arraystretch}{1.15}
\caption{CIFAR-100 results. Top-1 accuracy ($\%$) is adopted as the evaluation metric. \textcolor{red}{Red} values denote non-trivial improvement. \textcolor{blue}{Blue} values denote slight improvement less than $0.15$.}
\label{table_cifar}
\centering
\resizebox{\textwidth}{!}{
\begin{tabular}{c|cccccc|ccccc}
& \multicolumn{6}{c|}{Identical architecture structures} & \multicolumn{5}{c}{Distinct architecture structures}
\\
\hline
\multirow{2}{*}{Teacher}
& RN-56 & RN-110 & RN-110 & WRN-40-2 & WRN-40-2 & VGG-13 & WRN-40-2 & VGG-13 & RN-50 & RN-32$\times$4 & RN-32$\times$4
\\
& 72.34 & 74.31 & 74.31 & 75.61 & 75.61 & 74.64 & 75.61 & 74.64 & 79.34 & 79.42 & 79.42
\\
\multirow{2}{*}{Student}
& RN-20 & RN-32 & RN-20 & WRN-16-2 & WRN-40-1 & VGG-8 & SN-V1 & MN-V2 & MN-V2 & SN-V1 & SN-V2
\\
& 69.06 & 71.14 & 69.06 & 73.26 & 71.98 & 70.36 & 70.50 & 64.60 & 64.60 & 70.50 & 71.82
\\
\hline
KD \cite{kd} & 70.66 & 73.08 & 70.66 & 74.92 & 73.54 & 72.98 & 74.83 & 67.37 & 67.35 & 74.07 & 74.45
\\
Sort-KD & 71.22 & 73.71 & 71.08 & 75.10 & 74.32 & 73.54 & 75.72 & 68.37 & 68.50 & 74.43 & 75.34
\\
$\Delta$
& \textcolor{red}{(+0.56)} & \textcolor{red}{(+0.63)} & \textcolor{red}{(+0.42)} & \textcolor{red}{(+0.18)} & \textcolor{red}{(+0.78)} & \textcolor{red}{(+0.56)} & \textcolor{red}{(+0.89)} & \textcolor{red}{(+1.00)} & \textcolor{red}{(+1.15)} & \textcolor{red}{(+0.36)} & \textcolor{red}{(+0.89)}
\\
\hline
DKD \cite{dkd}
& 71.43 & 73.66 & 71.28 & 75.70 & 74.54 & 74.49 & 76.24 & 69.12 & 70.30 & 75.44 & 76.48
\\
Sort-DKD
& 71.62 & 73.96 & 71.67 & 75.95 & 74.62 & 74.73 & 76.31 & 69.83 & 70.42 & 76.15 & 77.04
\\
$\Delta$
& \textcolor{red}{(+0.19)} & \textcolor{red}{(+0.30)} & \textcolor{red}{(+0.39)} & \textcolor{red}{(+0.25)} & \textcolor{blue}{(+0.08)} & \textcolor{red}{(+0.24)} & \textcolor{blue}{(+0.07)} & \textcolor{red}{(+0.71)} & \textcolor{blue}{(+0.12)} & \textcolor{red}{(+0.71)} & \textcolor{red}{(+0.56)}
\\
\hline
CTKD \cite{ctkd}
& 71.19 & 73.52 & 70.99 & 75.45 & 73.93 & 73.52 & 75.78 & 68.46 & 68.47 & 74.48 & 75.31
\\
Sort-CTKD
& 71.41 & 73.90 & 71.28 & 75.53 & 74.44 & 73.84 & 76.15 & 68.61 & 68.54 & 74.57 & 75.67
\\
$\Delta$
& \textcolor{red}{(+0.22)} & \textcolor{red}{(+0.38)} & \textcolor{red}{(+0.29)} & \textcolor{blue}{(+0.08)} & \textcolor{red}{(+0.51)} & \textcolor{red}{(+0.32)} & \textcolor{red}{(+0.37)} & \textcolor{red}{(+0.15)} & \textcolor{blue}{(+0.07)} & \textcolor{blue}{(+0.09)} & \textcolor{red}{(+0.36)}
\\
\hline
LSKD \cite{lskd}
& 71.43 & 74.17 & 71.48 & 76.11 & 74.37 & 74.36 & 76.45 & 68.61 & 69.02 & 74.87 & 75.56
\\
Sort-LSKD
& 71.51 & 74.36 & 71.73 & 76.23 & 75.03 & 74.67 & 76.67 & 69.15 & 69.65 & 75.63 & 76.41
\\
$\Delta$
& \textcolor{blue}{(+0.08)} & \textcolor{red}{(+0.19)} & \textcolor{red}{(+0.25)} & \textcolor{blue}{(+0.12)} & \textcolor{red}{(+0.66)} & \textcolor{red}{(+0.31)} & \textcolor{red}{(+0.22)} & \textcolor{red}{(+0.54)} & \textcolor{red}{(+0.63)} & \textcolor{red}{(+0.76)} & \textcolor{red}{(+0.85)}
\\
\hline
\end{tabular}
}
\end{table*}

\begin{table*}[ht]
\renewcommand{\arraystretch}{1.15}
\caption{ImageNet results. Top-1 and top-5 accuracy ($\%$) are reported. \textcolor{red}{Red} values denote non-trivial improvement. \textcolor{blue}{Blue} values denote slight improvement less than $0.15$.}
\label{table_imagenet_res34_res18}
\centering
\resizebox{\textwidth}{!}{
\begin{tabular}{l|c|cc|cc|cc|cc}
Model & & teacher & student & KD & Sort-KD & DKD & Sort-DKD & LSKD & Sort-LSKD
\\
\hline
\multirow{2}{*}{ResNet34/ResNet18} & Top-1 & 73.31 & 69.75 & 70.87 & 71.18 \textcolor{red}{(+0.31)} & 71.70 & 71.84 \textcolor{blue}{(+0.14)} & 71.42 & 71.71 \textcolor{red}{(+0.29)}
\\
& Top-5 & 91.42 & 89.07 & 90.02 & 90.26 \textcolor{red}{(+0.23)} & 90.41 & 90.52 \textcolor{blue}{(+0.11)} & 90.29 & 90.55 \textcolor{red}{(+0.26)}
\\
\hline
\multirow{2}{*}{ResNet50/MN-V1} & Top-1 & 76.16 & 68.87 & 70.50 & 70.70 \textcolor{red}{(+0.20)} & 72.05 & 72.43 \textcolor{red}{(+0.38)} & 72.18 & 72.65 \textcolor{red}{(+0.47)}
\\
& Top-5 & 92.86 & 88.76 & 89.80 & 89.99 \textcolor{red}{(+0.19)} & 91.05 & 91.17 \textcolor{blue}{(+0.12)} & 90.80 & 91.22 \textcolor{red}{(+0.42)}
\\
\end{tabular}
}
\end{table*}

\subsection{Implementation Detail}
All experiment results are averaged over four runs. For CIFAR-100, we train the models for 240 epochs with 64 batch size. We follow \cite{crd} training settings. The initial learning rate is 0.01 for MobileNets and ShuffleNets, and 0.05 for other architecture types (\textit{e.g.}, VGGs, ResNets, and WRNs). The learning rates decay by 0.1 at the 90th, 180th, and 210th epochs. We use SGD with 0.9 momentum and 5e-4 weight decay as the optimizer. For ImageNet, all models are trained for 100 epochs with a 512 batch size. The initial learning rate is 0.2. The learning rates decay by 0.1 at the 30th, 60th, and 90th epochs. We use SGD with 0.9 momentum and 1e-4 weight decay as the optimizer.

\begin{table}[h]
\renewcommand{\arraystretch}{1.15}
\caption{Comparison with a swap method.}
\label{table_comparison}
\centering
\resizebox{\columnwidth}{!}{\begin{tabular}{c|cccc|c}
Dataset & \multicolumn{4}{c|}{CIFAR-100} & ImageNet
\\
\hline
\multirow{2}{*}{Teacher} & RN-110 & WRN-40-2 & VGG-13 & RN-32x4 & RN-34
\\
& 74.31 & 75.61 & 74.64 & 79.42 & 73.31
\\
\multirow{2}{*}{Student} & RN-32 & WRN-40-1 & MN-V2 & SN-V2 & RN-18
\\
 & 71.14 & 71.98 & 64.60 & 71.82 & 69.75
\\
\hline
KD  & 73.08 & 73.54 & 67.37 & 74.45 & 70.87
\\
w/ Swap & 73.35 & 74.07 & 67.85 & 74.93 & 70.92
\\
w/ Sort (ours) & \textbf{73.71} & \textbf{74.32} &  \textbf{68.37} & \textbf{75.34} & \textbf{71.18}
\\
\end{tabular}}
\end{table}

\subsection{Experimental Results}
\noindent\textbf{Results on CIFAR-100.}
Table \ref{table_cifar} shows the top-1 image classification accuracy on CIFAR-100 with various teacher-student pairs. As a plug-and-play processing, we evaluate our method on four existing logit-based KD methods, such as KD, DKD, CTKD, and LSKD. We show that all student models benefit from our sorting mechanism, and the improvement is quite significant in some cases. Additionally, we also provide a noisy label \cite{noisylabel} experiment in the Appendix.

\noindent\textbf{Results on ImageNet.}
Top-1 and top-5 accuracies of image classification on ImageNet are reported in Table \ref{table_imagenet_res34_res18}. We apply our sorting mechanism to KD, DKD, and LSKD. As a result, our method can achieve consistent improvements in logit-based KD methods. Particularly, it demonstrates improvements not only for top-1 but also for top-5 accuracy. This is due to the modification of the teacher's top prediction confidence on misclassified samples through a sorting mechanism.

\noindent\textbf{Comparison with Swap Method.}
Our proposed sorting mechanism is motivated by the swap method and serves as a refined modification of this method. In Table \ref{table_comparison}, we show that the sorting mechanism consistently outperforms the swap method on CIFAR-100 and ImageNet datasets.

\section{Conclusion}
In this letter, we revisit the traditional logit-based knowledge distillation method and highlight that directly using the teacher's outputs makes the student inferior on misclassified samples. To address this issue, we propose a sorting mechanism to modify the teacher's prediction based on label information. Concretely, the target confidence is adjusted to be the highest, and the rest of the distribution is reordered by confidence ranking. Compared with the swap method, our sorting mechanism does not drastically devalue the highly correlated semantics but distributes them by ranking. As a result, sort-KD can achieve better results because it is natural, robust, and efficient. The extensive experiments on several benchmark datasets demonstrate the effectiveness of our proposed method in improving the existing logit-based KD methods across a range of teacher-student pairs.

\bibliographystyle{IEEEtran}
\bibliography{sort}

\clearpage 

\setcounter{table}{0}
\renewcommand{\thetable}{A\arabic{table}}

\setcounter{figure}{0}
\renewcommand{\thefigure}{A\arabic{figure}}

\appendices

\section{Supplementary}

\begin{table}[ht]
\renewcommand{\arraystretch}{1.1}
\caption{Performance on MS-COCO based on Faster-RCNN \& FPN. AP is the evaluation metric. \textcolor{red}{Red} values denote non-trivial improvement.}
\label{table_objectdetection}
\centering
\resizebox{\columnwidth}{!}{
\begin{tabular}{c|cccccc}
 & AP & AP50 & AP75 & APl & APm & APs
\\
\hline
T: R-101 & 42.04 & 62.48 & 45.88 & 54.60 & 45.55 & 25.22
\\
S: R-18 & 33.26 & 53.61 & 35.26 & 43.16 & 35.68 & 18.96
\\
\hline
KD & 33.97 & 54.66 & 36.62 & 44.14 & 36.67 & 18.71
\\
Sort-KD & 34.38 & 55.38 & 36.90 & 45.04 & 36.84 & 19.29 
\\
$\Delta$ & \color{red}(+0.41) & \color{red}(+0.72) & \color{red}(+0.28) & \color{red}(+0.90) & \color{red}(+0.17) & \color{red}(+0.58) 
\\
\hline \hline
T: R-50 & 40.22 & 61.02 & 43.81 & 51.98 & 43.53 & 24.16
\\
S: MV-2 & 29.47 & 48.87 & 30.90 & 38.86 & 30.77 & 16.33
\\
\hline
KD & 30.13 & 50.28 & 31.35 & 39.56 & 31.91 & 16.69
\\
Sort-KD & 31.33 & 52.46 & 32.69 & 41.02 & 33.59 & 18.19
\\
$\Delta$ & \color{red}(+1.20) & \color{red}(+2.18) & \color{red}(+1.34) & \color{red}(+1.46) & \color{red}(+1.68) & \color{red}(+1.50)
\\
\hline
\end{tabular}
}
\end{table}

\noindent\textbf{Results on MS-COCO.}
We apply our sorting mechanism to classical KD in the object detection task. MS-COCO (2017) is a standard object detection dataset with 80 classes. The train split contains 118,000 images, and the validation split contains 5,000 images. We follow the implementation of DKD for object detection. As shown in Table \ref{table_objectdetection}, our sorting mechanism can boost the classical KD detection performance. This demonstrates that the improvement is not restricted to image classification tasks.

\begin{figure}[h]
\includegraphics[scale=0.4]{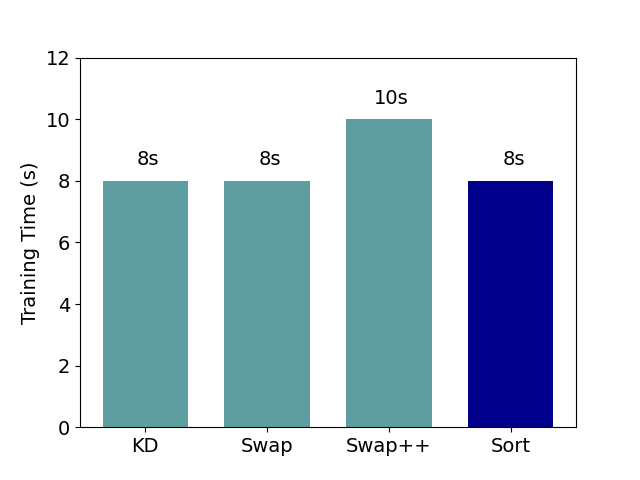}
\centering
\caption{Training time in seconds (per epoch). We set ResNet110 as the teacher and ResNet32 as the student on CIFAR-100.}
\label{fig:time}
\end{figure}

\begin{figure}[h]
\includegraphics[scale=0.24]{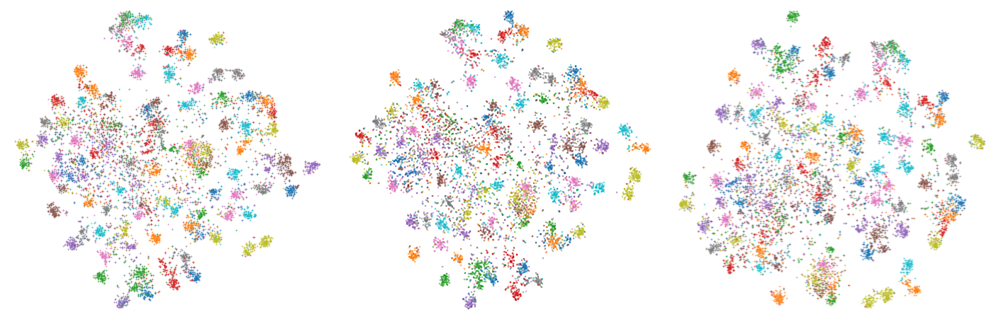}
\centering
\caption{t-SNE of features representation by classical KD (left), Swap (center), and Sort-KD (right). The visualizations are taken from RN-110/RN-32 teacher-student pairs on CIFAR-100.}
\label{fig:tsne}
\end{figure}

\noindent\textbf{Training Efficiency.}
We assess the training time to evaluate the efficiency of our sorting mechanism. As shown in Fig. \ref{fig:time}, the training time per epoch is the same as the classical KD and swap method. We also demonstrate the training cost of swap++, which is a swap method that is applied multiple times to produce results equivalent to those generated by our sorting mechanism. We observe that our sorting mechanism can cut down the training cost, demonstrating that the sorting mechanism is efficient.

\noindent\textbf{Feature Visualization.}
In Fig. \ref{fig:tsne}, we visualize the deep representation of the student model. It shows that the representation of KD with a sorting mechanism is more separable than the default and swap method, showing the discriminability of students. Specifically, the overall representation is a circle-like form, while the other methods are rhombus-like shapes.

\begin{figure}[h]
\includegraphics[scale=0.34]{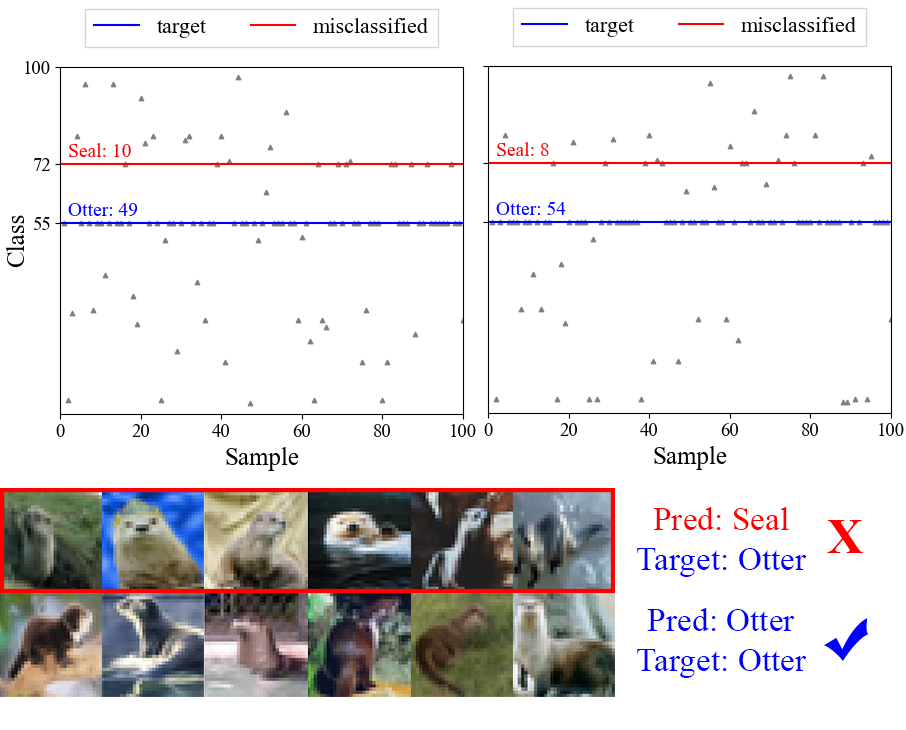}
\centering
\caption{Prediction of classical KD (left) and Sort-KD (right) on CIFAR-100. Class “55” (otter) is the target, and class “72” (seal) is misclassified. The total number of samples is 100. The results are taken from RN-110/RN-32 teacher-student pairs. Both targets and misclassified images are shown at the bottom.}
\label{fig:dist}
\end{figure}

\noindent\textbf{Prediction Analysis.}
As shown in Fig. \ref{fig:dist}, we verify the effectiveness of Sort-KD compared with the classical KD through predictions on test samples. Specifically, both the otter as a target and the seal share similar semantics as they belong to the same superclass of aquatic mammals. In this case, the student could be confused when receiving the misclassified prediction from the teacher. Our sorting mechanism can tackle this issue by refining the prediction via a label. To this end, we demonstrate that Sort-KD prediction is better than classical KD's.

\begin{table}[h]
\renewcommand{\arraystretch}{1.1}
\caption{Top-1 on CIFAR-100 training set with different noisy ratios. We take RN-110 as teacher and RN-32 as student.}
\label{table_noise}
\centering
\begin{tabular}{c|c|c|c}
noisy ratio & Sort & Top-1 ($\%$) & $\Delta$
\\
\hline
\multirow{2}{*}{0.1} &  & 65.83 & -
\\
 & \cmark & 66.04 & +0.21
\\
\hline
\multirow{2}{*}{0.2} &  & 65.45 & -
\\
 & \cmark & 65.48 & +0.03
\\
\hline
\multirow{2}{*}{0.3} &  & 65.08 & -
\\
 & \cmark & 65.46 & +0.38
\\
\end{tabular}
\end{table}

\noindent\textbf{Noisy Label.}
In Table \ref{table_noise}, we evaluate our method on CIFAR-100 with [0.1, 0.2, 0.3] symmetric noisy ratios. The results suggest that the sorting method demonstrates enhanced performance when applied to training data with higher levels of noise.

\end{document}